# Magnetic spin-orbit interaction directs Bloch surface waves

Mengjia Wang, † Hongyi Zhang, † Tatiana Kovalevitch, † Roland Salut, † Myun-Sik Kim, $ Miguel Angel Suarez, † Maria-Pilar Bernal, † Hans-Peter Herzig, $ Huihui Lu, ¶§ and Thierry Grosjean, †*

† FEMTO-ST Institute, Université Bourgogne Franche-Comté, UMR CNRS 6174 15B Av. des Montboucons, 25030 Besancon cedex, France
$ Optics & Photonics Technology Laboratory, Ecole Polytechnique Fédérale de Lausanne (EPFL), Rue de la Maladière 71b, Neuchâtel, CH-2000, Switzerland
¶ Key Laboratory of Optoelectronic Information and Sensing Technologies of Guangdong Higher Educational Institutes, Jinan University,Guangzhou, China
§email: thuihuilu@jnu.edu.cn

* E-mail: thierry.grosjean@univ-fcomte.fr,
Phone: +33 (0)3 81 66 64 17. Fax: +33 (0)3 81 66 64 23

We study the directional excitation of optical surface waves controlled by the magnetic field of light. We theoretically predict that a spinning magnetic dipole develops a tunable unidirectional coupling of light to TE-polarized Bloch surface waves (BSWs). Experimentally, we show that the helicity of light projected onto a subwavelength groove milled in the top layer of a 1D photonic crystal (PC) controls the power distribution between two TE-polarized BSWs excited on both sides of the groove. Such a phenomenon is shown to be mediated solely by the helicity of the magnetic field of light, thus revealing a magnetic spin-orbit interaction. Remarkably, this magnetic optical effect is clearly observed with a near-field coupler governed by an electric dipole moment: it is of the same order of magnitude as the electric optical effects involved in the coupling. It opens new degrees of freedom in the manipulation of light and offers appealing novel opportunities in the development of integrated optical functionalities.

## INTRODUCTION

The magnetic field of light is often considered to be a negligible contributor to light-matter interaction. However, with the advent of the left-handed metamaterials[1–4], nanophotonics has recently investigated magnetic response in nanostructures to reveal the hidden magnetic part of the light-matter interaction, e.g., to achieve negative refractive indices[4], to control magnetic transitions in matter[5–7], to map optical magnetic fields[8–13] and to study magnetic effects at optical frequencies[14–16]. In this paper, we show that the magnetic field of light has also the appealing ability to control the light coupling into optical surface waves.

Optical angular momenta are manifestations of polarization and spatial degrees of freedom of light[17]. Remarkably, spin and orbital momenta are not independent quantities, the spin angular momentum (SAM) can be converted into orbital angular momentum (OAM), and vice versa[18]. Such a spin-orbit interaction (SOI) has recently drawn much interest for applications involving light manipulation[19–25]. For example, SOI has demonstrated the remarkable property of controlling the propagation direction of guided modes, such as surface plasmons and fiber modes, leading to the concept of a spin-controlled unidirectional waveguiding[19,26–32]. A robust spin-controlled unidirectional waveguiding relies on the transverse SAM arising in evanescent waves[33-36]. With the help of a subwavelength (dipolar) coupler, the longitudinal SAM of an impinging wave can be transferred into the transverse SAM of the evanescent tail of a guided mode, leading to a spin-directional coupling of the guided mode[19,28,29]. So far, such investigations focused on the rotating electric component of light as the source of SAM originating the transverse spin-direction coupling[34].

Here, we introduce the concept of a spin-direction locking mediated solely by a rotating magnetic light field. We study the light coupling in Bloch surface waves (BSW) by projecting circularly polarized light onto a subwavelength scatterer used as a near-field coupler. BSWs are pure evanescent modes bound to the free surface of a 1D photonic crystal[37,38]. Importantly, when the BSW is TE-polarized, it is described by a rotating magnetic field (Fig. 1 (b)), instead of a rotating electric field as for the (TM-polarized) surface plasmons or the guided mode of a nanofiber[18,34]. We show numerically that the tunable unidirectional excitation of TE-polarized BSWs is achievable with a spinning magnetic dipole source, thus proving that the rotating magnetic field of a BSW carries SAM. Using a subwavelength groove as a light-to-BSW converter, we observe that the directionality of the incoupled light is helicity dependent. From the intrinsic spin properties of TE-polarized evanescent waves and an analytical model of the coupling, we infer that the resulting spin-controlled directional coupling is mediated by the magnetic field of light, thus revealing a magnetic SOI. Despite the electric dipole nature of the groove, this magnetic optical effect is found to be of the order of magnitude of the electric effects involved in the coupling process.

**METHODS**

**Numerical models**
All the numerical simulations have been carried out with the finite difference time domain method (FDTD, commercial code Fullwave).

The model used for the calculations, by 2D FDTD, of the dipole-to-BSW near-field coupling consists of an area spanning ±9 μm along the $x$ direction and ±5.5 μm along the $z$ direction about the dipole. The system is invariant along the $y$ direction. The 1D photonic crystal design is described below. All four boundaries of the computation volume are terminated with perfectly matched layers in order to avoid parasitic unphysical reflections around the structure. The grid resolution is 10 nm. For each type of dipole (electric dipole (ED) and magnetic dipole (MD)), two simulations are realized with the dipole oriented along $x$ and $z$ axis, respectively. In both cases, the electromagnetic fields across the structure are recorded. Then, the BSW excitations with spinning ED and MD are reconstructed from these two sets of simulations. In both cases, the helicity of the dipole moments are defined analytically to match the helicity of the magnetic field of the BSWs under study.

The model used for the calculation, by 3D FDTD, of the light/BSW coupling with a single groove consists of a volume which spans ± 5 μm along both the $x$ and $y$ directions perpendicular and parallel to the groove, respectively. The groove, with a width and a depth of 600 nm, is engraved on the top layer of the 1D photonic crystal considered in this study. It is located at $x = z = 0$ and extends in the $y$ direction across the computation volume. The simulation spans 4.9 μm below the 1D photonic crystal top surface in the glass substrate and terminates 2 μm in air, beyond the multilayer. The operating wavelength is 1.55 μm. All six boundaries of the computation volume are terminated with perfectly matched layers in order to avoid parasitic unphysical reflections around the structure. The non-uniform grid resolution varies from 20 nm for portions at the periphery of the simulation to 8 nm within and near the top layer. In order to excite the BSWs, a Gaussian beam is launched onto the groove. Its propagation axis is tilted by 80° with respect to the surface normal. Due to computation limitations, a beam waist of 1.5 μm has been chosen in order to limit the computation volume. Two simulations are conducted with incident beams of TE and TM polarizations with respect to the surface. In both cases, the intensities of the excited BSWs on the right and left sides of the groove are recorded. Afterward, the various BSW excitation scenarios are reconstructed from these two simulations. To this end, the various input polarizations are expressed in the TE/TM local coordinate frame of the incidence beam.

**Experimental illumination system**
The illumination system consists of two lenses coupled to a polarizer and a quarter-wave plate (QWP), (see details in Supplementary Material and Supplementary Fig. S1). The first lens ($f = 33$ mm) collimates the light emerging from a laser source coupled into a single mode fiber, whereas the second lens ($f$ = 50 mm), positioned closer to the sample, focuses the collimated beam onto the groove. The polarizer and QWP are positioned between the two lenses. The polarizer ensures a TM linearly polarized incident beam with respect to the sample surface. It is followed by the QWP for manipulating the polarization ellipticity of the incident beam. By rotating the QWP, the polarization of the incident beam is tuned from linear to circular polarization, with intermediate elliptic polarization states. Importantly, in this configuration, the orientation of the polarization ellipse varies with the orientation of the quarter-wave plate. A linear combination of the BSW intensities is then performed with the projection amplitudes as the coefficients of the linear combination.

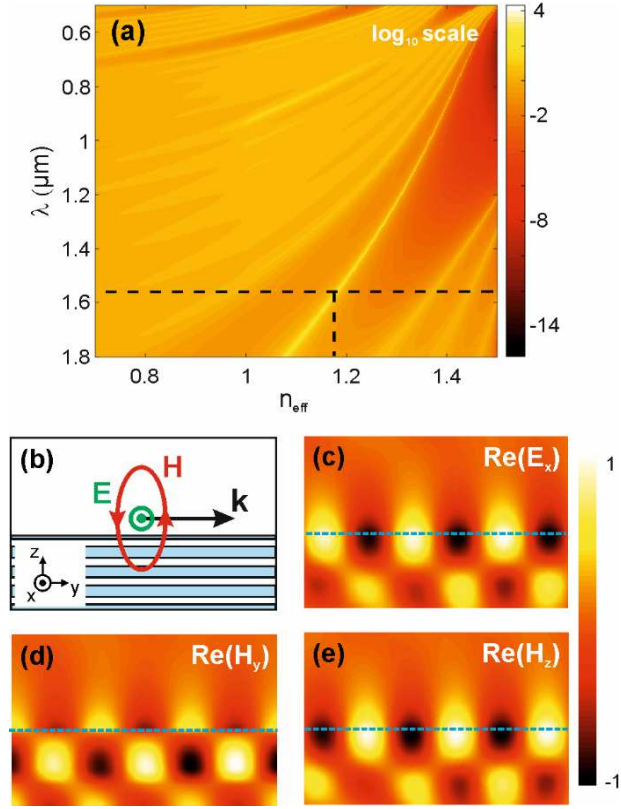

**Figure 1:** (a) Dispersion diagram of the 1D photonic crystal with a log-scale colorbar. The structure generates a photonic bandgap at the middle of which the dispersion curve of a BSW is observed. (b) Schematic diagram of the electromagnetic field distribution of a TE-polarized BSW. (c) Simulation result of the real part of the electric field component $E_x$ parallel to the sample surface. (d) and (e) Simulation results of the real part of $H_y$ and $H_z$, respectively.

**RESULTS AND DISCUSSIONS**

BSWs are generated by a 1D photonic crystal consisting of a stack of 6 pairs of silicon dioxide and silicon nitride layers, with refractive indices of 1.45 and 1.79 (at $\lambda$ = 1.55 µm), and thicknesses of 492 nm and 263 nm, respectively. The multilayer lies on a glass substrate (refractive index of 1.5) and it is covered with a thin 80-nm-thick layer of silicon nitride. Fig. 1(a) represents the dispersion diagram of the structure (calculation by the impedance approach[39]). Such a diagram shows a photonic bandgap, which contains the dispersion curve of a BSW. Given the 1D PC design, this surface mode is TE-polarized, as shown in Fig. 1(b-d). We observe that $H_x$ (Fig. 1(c)) is shifted by a quarter wavelength with respect to $H_z$ (Fig. 1(c)) along the propagation direction $x$ of the surface wave, thus revealing the helicity of the optical magnetic field along the transverse $y$ direction. The transverse SAM of the surface wave is thus carried solely by its rotating magnetic field, the electric field shows no helicity.

We numerically study the coupling of single electric and magnetic dipoles to a TE-polarized BSW. To this end, the dipoles are considered to be positioned 10 nm above the top surface of the 1D photonic crystal described above, which radiate light in continuous wave regime at $\lambda = 1.55$ µm (details in Materials and Methods section). The electric dipole (ED) is oriented along the $y$ direction along the surface, i.e., parallel to the electric field of the TE-polarized BSW. The magnetic dipole (MD) rotates in the ($xz$) plane perpendicular to the surface, i.e., in the helicity plane of the rotating magnetic field of the TE-polarized BSW. Fig. 2 shows a snapshot of the resulting electric field amplitude along the ($xz$) plane, for the ED (Fig. 2(a)) and MD (Fig. 2(b) and (c)) excitations. The MD, whose dipole moment is $\vec{m} \propto \vec{e}_x \pm j * 0.53 \vec{e}_z$ ($j = \sqrt{-1}$), rotates either anti-clockwise (Fig. 2(b)) or clockwise (Fig. 2(c)). In these figures, the field distributions around the dipoles are saturated in order to provide a better view of the light distributions at the structure surface. The simulations are carried out with the 2D FDTD method. With the ED, the BSW is symmetrically excited on both sides of the point-like source. No directionality is observed in the optical coupling process. In contrast, the optical coupling process becomes unidirectional with the spinning MD. The portion of the incoupled energy that propagates in one of the two possible directions is larger than 99% of the total incoupled energy (see Fig. 2(d)). Moreover, the propagation direction of the BSW is controlled by rotating the direction of the MD source, as shown in Figs. 2(b)-(d). These results reveal a tunable unidirectional optical coupling controlled by the magnetic field of light. They also confirm that the rotating magnetic field of a TE-polarized BSW carries SAM.

From an experimental point-of-view, the achievement of a tunable directional coupling of BSW with a spinning MD would be possible using a dielectric sphere showing magnetic resonances[6,40,41] directly deposited on the top of the 1D photonic crystal. The bead would then be illuminated with a circularly polarized beam at near-grazing incidence, following the electric spin-controlled excitation process of surface plasmons[29]. We will show here that this magnetic directional coupling can be clearly evidenced even with a standard subwavelength groove directly engraved on the top of the 1D photonic crystal.

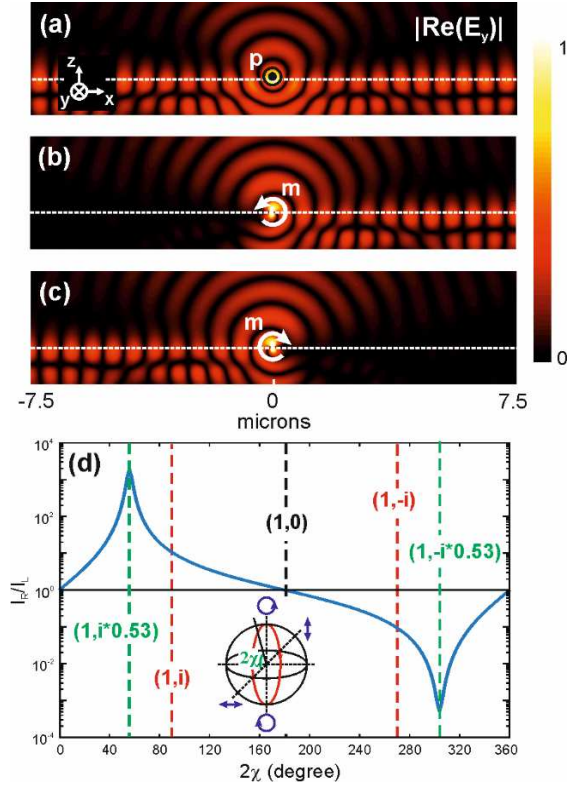

**Figure 2:** Simulation by FDTD of the coupling of (a) an ED source oriented along $y$ axis, and (b) and (c) a spinning MD source, to a TE-polarized BSW. The absolute value of the real part of the electric field is represented in false colors. The MD rotates either (b) anti-clockwise or (c) clockwise. (d) Directionality factor (ratio between electric intensities of the left and right BSWs) for various MD polarizations. The MD polarization ellipticity is changed along the path shown in the Poincare sphere (see inset). The MD moment is also expressed at specific values of the ellipticity angle $2\chi$.

In order to produce the above described 1D photonic crystal, thin layers of silicon oxide and silicon nitride are deposited alternately by plasma enhanced chemical vapor deposition (PECVD) onto a glass wafer. A cross-section of the multilayer realized by focused ion beam (FIB) reveals the design detailed above (Fig. 3(a)). Then, the sample is covered by a 100-nm thick chromium layer and a 600-nm wide and deep groove is milled by FIB over a length of 20 µm. Finally, chromium is removed. The inset of Fig. 3(b) shows a scanning electron microscope (SEM) image of a resulting structure.

The structure is characterized in the far-field by projecting a slightly focused beam of controlled polarization onto the sub- wavelength groove, at an incidence angle of about 80°. The structure is imaged in reflection mode with an objective (20X, NA = 0.4) coupled to an infrared camera (see details in the Supplementary Material and Supplementary Fig. S1). Owing to the light scattering to the free surface of the 1D photonic crystal, a direct real-time mapping of the surfaces waves excited on both sides of the groove is possible. Fig. 3(b) shows the far-field images of the surface around the groove under illumination. The incident beam is TM linearly polarized with respect to the sample surface, leading to a symmetric scattering pattern. Note that the definition of the TE/TM polarizations for the incident light and the BSWs are related to different local coordinate frames, and thus they should not be directly compared. In the context of our study, a TM-polarized incident wave can excite a TE-polarized BSW. The bright elongated spot along the $y$-axis is the cross-section of the excitation beam along the surface. The two narrow rays on both sides of the excitation spot are the traces of the BSWs excited by the subwavelength groove. Linear momentum conservation imposes a tilt angle of the BSW propagation direction with respect to the groove direction ($y$) predicted to be about 33.8°, and measured to a value around 36°.

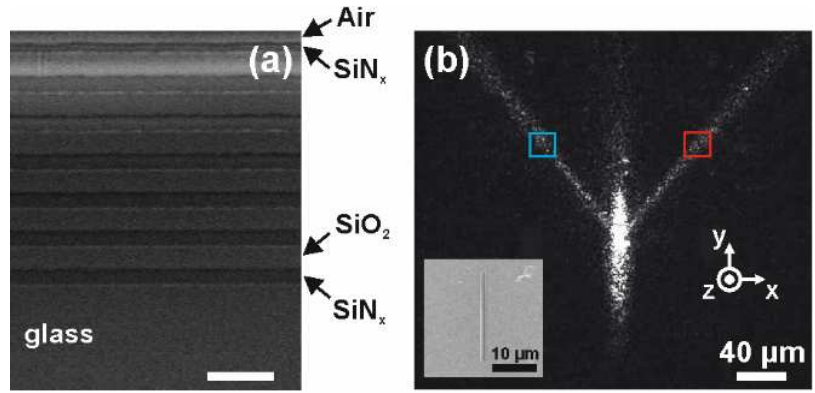

**Figure 3:** (a) SEM cross section of the 1D photonic crystal (scale bar: 1 µm). (b) Far-field optical image of the BSW obtained with an excitation at $\lambda = 1.55$ µm of a 600-nm wide and deep groove. The laser beam is incident from air on the top surface of the 1D PC at almost grazing angle (incidence angle: 80°, see Supplementary Fig. S1). This image originates from light scattering at the sample top surface. Figure inset: SEM top view of the groove milled into the top surface of the 1D photonic crystal.

We then study the distribution of incoupled power between the two surface waves, as a function of the incident polarization. To this end, images of the structure are acquired while varying the polarization of the incident beam. The polarization is defined by the angle $\theta$ between the axes of the quarter-wave plate and the polarizer When $\theta = k\,90°, k = 0, 1, 2, 3$, the polarization is linear whereas a circular polarization is achieved for $\theta = 45° + k\,90°$, $k = 0, 1, 2, 3$. For intermediate angles, the polarization is elliptical with a major axis orientation defined by $\theta$. For each image recorded at a specific polarization state, we integrate the signal detected over two square areas located symmetrically with respect to the groove (shown in light red and blue colors in Fig. 3(b)). Finally, the resulting values $S_r$ and $S_l$ measured on the right and left BSWs, respectively, are plotted as a function of the angle $\theta$ (Fig. 4(a)). The experimental plots are represented in solid lines together with the simulation results obtained with the 3D FDTD method (see details in the Material and Methods section).

Fig. 4(a) reports $S_r$ and $S_l$ as a function of $\theta$. $S_r$ and $S_l$ are described by sinusoidal functions, shifted by about 30° from each other, whose amplitudes are modulated by a sinusoidal function. The experimental results and simulation predictions are in good agreement. As expected, the coupling process of light in the BSWs is asymmetric excepted for the linear (TM) incident polarization ($\theta = k90°, k = 0, 1, 2, 3$). In that case, the optical system is fully symmetric with respect to the groove and the two curves of $S_r$ and $S_l$ merge. By Fourier transforming these two functions (cf. Figs. 4(b), (c)), we see that they can be simply expressed analytically as:

$$S_i(\theta) = A_i^{(0)} + A_i^{(2)} \sin(2\theta + \phi_i^{(2)}) + A_i^{(4)} \sin(4\theta + \phi_i^{(4)}), \tag{1}$$

where $i = r, l$, and coefficients $A_i^{(u)}$ and $\phi_i^{(u)}$ ($u$=0, 2, 4) are constant. Coefficients $A_i^{(u)}$ and $\phi_i^{(u)}$ ($u$=2, 4) are given by the Fourier transform of $S_r$ and $S_l$.

We see in Fig. 4 (d) that the second harmonic components relative to the left and right BSWs are almost in opposition, i.e., shifted by 180°. The fourth harmonics (cf. Fig. 4 (e)) undergo the shift of about 30° initially evidenced in Fig. 4 (a). Importantly, the local maxima and minima of the second harmonic component closely coincide with the right and left-handed circular polarization states. Moreover, changing incident polarization handedness inverts the distribution of incoupled light in the right and left BSWs. The second harmonic contribution to the BSW coupling is therefore helicity dependent. In contrast, the fourth harmonics of the Fourier series stay unchanged when the input polarization handedness is reversed (See Figs. 4 (e)). Therefore, the fourth harmonic contribution to the BSW coupling is independent of the helicity of light.

Because subwavelength scatterers are optically governed by an electric dipole moment, one may consider that a subwavelength groove on top a 1D photonic crystal interacts with the electric field of an incoming wave to transfer energy to the BSWs. Following this pure electric model, and assuming an incident plane wave, coefficients $S_r$ and $S_l$ defined previously become proportional to the coupling rates:

$$R_i = \alpha \left| \vec{e}_i \cdot \vec{E}_{inc}(\vec{r}_0) \right|^2, \tag{2}$$

where $i = r, l$ denotes the right and left sides of the groove and $\alpha$ is a constant. $\vec{E}_{inc}(\vec{r}_0)$ is the incident electric field at a single point of coordinate $\vec{r}_0$ along the subwavelength groove. $\vec{e}_i$ is the unit vector in the directions of the electric field of the emerging right and left BSWs. When plotted as a function of the angle $\theta$, $R_r$ and $R_l$ are described by two sinusoids showing $4\theta$ dependence, and shifted by 8° (see Supplementary Material and Supplementary Fig. S3). Moreover, changing the polarization handedness does not interchange the values of the two coefficients, which evidences that a pure electric coupling of an incident light to the TE-polarized surface waves is helicity independent. By comparing the Supplementary Fig. S3 and Fig. 4(e), we see that $R_r$ and $R_l$ closely match the fourth harmonic function of Eq. 1. The unbalanced electric coupling is due to the asymmetric projections of the electric field onto $\vec{e}_r$ and $\vec{e}_l$. The larger angular shift observed between the couple of experimental curves (30° versus 10° with our model) may be due to the fact that the scattering properties of our 600 nm large (i.e., $\lambda/2.5$) groove-like coupler slightly deviates from the dipole emission.

The modulation of this electric coupling by a helicity dependent optical process (Fig. 4) is not predicted by our analytical model. As noted above, only the magnetic field of the TE polarized BSW is rotating, the electric field shows zero helicity. An helicity dependent process on such waves thus solely involves the magnetic field of light. We plotted, as a function of $\theta$, the ellipticity factor of the magnetic field (plane-wave illumination) projected in the helicity planes of the right and left BSWs (i.e., the planes perpendicular to the transverse spin momentum of the surface waves). Details of the calculation are given in the Supplementary Material. The ellipticity curves show a periodicity of $2\theta$ and lead to opposite values when the polarization handedness is changed (see Supplementary Fig. S4). These curves closely resemble the second harmonic functions of Eq. 1 (Figs. 4 (d)). Therefore, the second harmonic contribution to the optical coupling is solely controlled by the magnetic field of light.

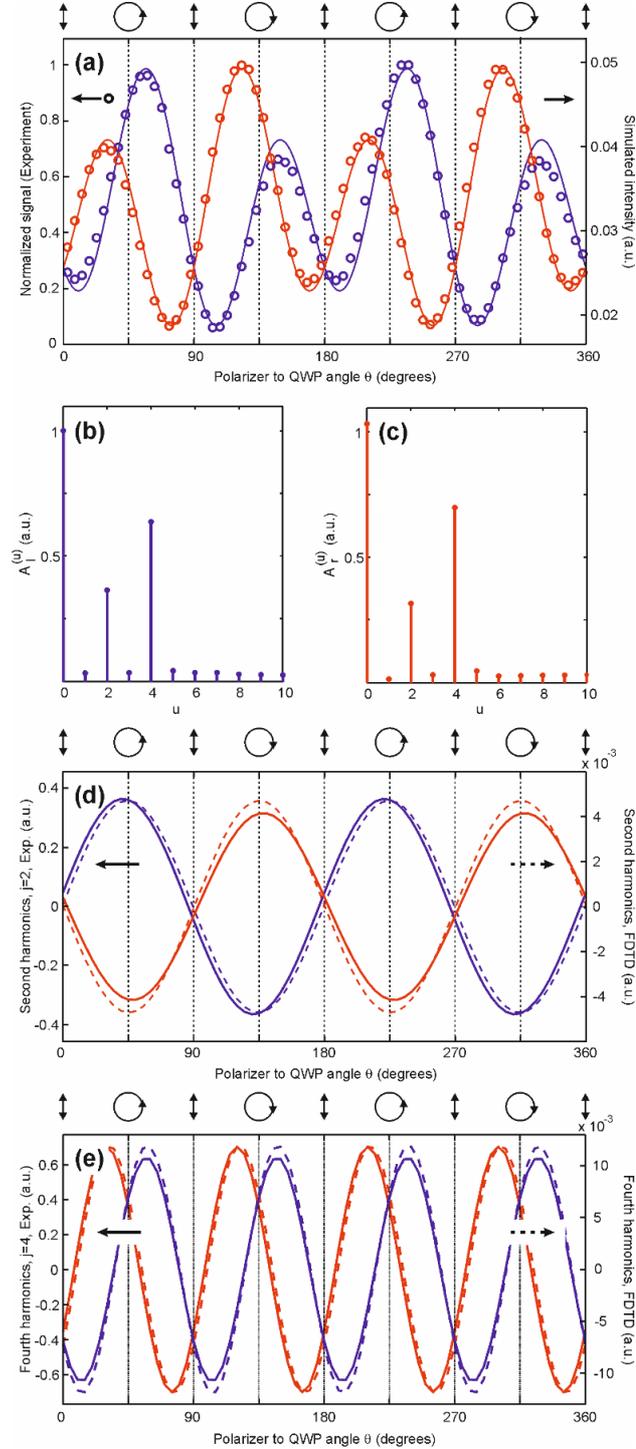

**Figure 4:** (a) Detected signals (circles) and simulated intensities (FDTD method, dashed lines) on the right and left BSWs, as a function of the angle $\theta$ between the quarter-wave plate and the polarizer. The curves related to the right and left BSWs are represented in blue and red colors, respectively. (b) and (c) Spectrum (amplitude) of the experimental blue and red curves of (a), obtain by Fourier transform. Coefficient *s* defines the harmonic orders of the Fourier series. (d)

and (e) Representation in the real space of the non-null harmonics of the two Fourier series shown in (b) and (c): (d) second harmonics and (e) fourth harmonics. Experimental and numerical curves are plotted in solid and dashed lines, respectively.

Remarkably, this magnetic effect is clearly visible with a dielectric scatterer described by an electric dipole moment. Fig. 4 (b) and (c) show that its contribution is larger than 45% of the electric contribution to the coupling. Despite the extremely low response of the scatterer to the magnetic field of the impinging wave, the rotating magnetic field right at an electric scatterer provides the initial conditions to direct a large portion of the incoupled energy to the right or to the left BSW, depending on the polarization handedness. The second harmonic curves shown in Fig. 4(d) thus describe a magnetic spin-directional coupling, as evidenced in Fig. 2. In the experimental case however, the phase matching between the incident light and the BSW is mediated by the electric optical field, given the electric dipole nature of the scatterer. The rotating incident magnetic field right at the groove, which is almost not affected by the scatterer, controls the directionality of the launched surface waves. This explains the $4\theta$ dependence of the experimental coupling process (Fig. 4) that is comparatively not observed with the MD excitation: Fig. 2(c) shows a directionality curve with only a 180° periodicity. The BSW excitation undergoes the light-to-BSW electric field projection rules that come along with the phase matching. A subwavelength resonant particle or antenna, whose resonance is described by a MD moment, would cancel this electric component of the coupling. Such a configuration is however beyond the scope of this paper.

**CONCLUSION**

We have described a new magnetic effect in light-matter interaction, called magnetic SOI. On the basis of this magnetic SOI, we have shown that an elliptically polarized MD develops a tunable unidirectional coupling of light into TE-polarized BSWs: depending on MD's helicity, the surface waves propagate upstream or downstream. The underlying phenomenon is a transverse spin-direction coupling in BSWs, but here with the spin momentum described solely by a rotating magnetic light field. This phenomenon has been demonstrated with a simple subwavelength groove used as a light-to-BSW converter. Despite the electric dipole nature of this coupler, the magnetic effect is of the same order of magnitude as the electric effects in the coupling process. In that particular case, the magnetic field is not involved in the light-to-BSW phase matching process originating the transfer of energy to the BSW, but it controls the directionality of the incoupled energy. By using couplers developing magnetic resonances[6,40,41], a pure magnetic tunable unidirectional coupling would be possible experimentally, in accordance with our theoretically predictions. Besides the fundamental questions raised about the magnetic optical control of light-matter interaction, our results greatly open the possibilities of controlling the optical flows in ultracompact architectures. Reciprocally, BSWs can also be used as a probe to locally investigate the magnetic polarization properties of light.

**ACKNOWLEDGEMENT**

This work is funded by the SMYLE program. It has been realized in the context of the Labex ACTION program (contract ANR-11-LABX-01-01). It is supported by the French RENATECH network and its FEMTO-ST technological facility.

# Supplementary Information

### 1- Experimental setup

The schematic diagram of the experimental setup is represented in Fig. S1. Light at λ=1.55 µm emerges from a tunable laser source (Agilent) coupled to a single mode fiber (SMF-28, Corning). It is collimated by the first lens ($f$ = 33 mm, Thorlabs) and focused by the second lens ($f$ = 50 mm, Thorlabs). The polarization of the collimated wave is manipulated by using a polarizer (LP, Thorlabs) and a quarter-wave plate (QWP, Thorlabs) positioned between the two lenses. The quarter-wave plate can be rotated at will with respect to the polarizer. Focused waves are projected directly on the groove at almost grazing incidence (incidence angle $\beta=80°$). The groove and its surrounding region covering the BSW propagation are imaged with a objective (20X, NA = 0.4) coupled to an infrared camera (GoldEye model G-033, Allied Vision Technologies GmbH). Far field imaging of the surface waves is rendered possible by the slight surface imperfection of the top layer of the 1D photonic crystal, which scatters the BSWs into the free space.

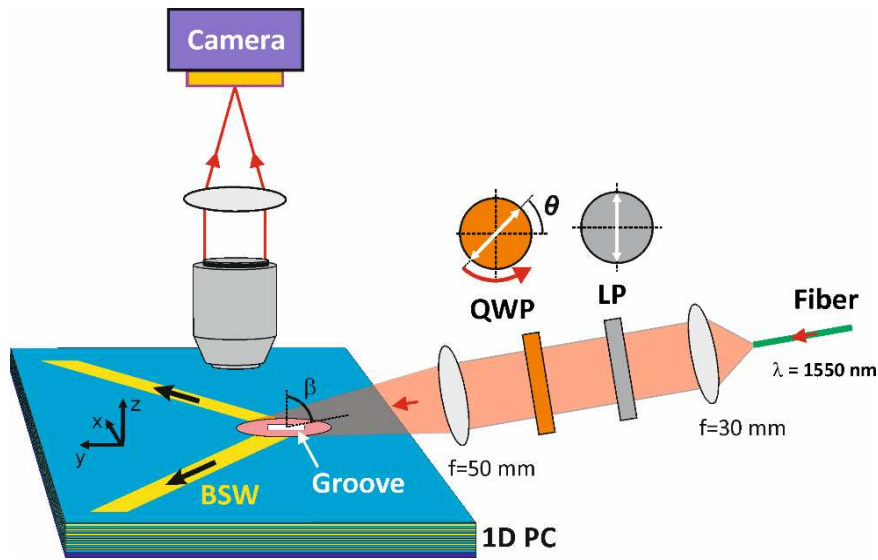

*Fig. S1. Schematic diagram of the experimental setup. BSW: Bloch surface wave, PC: photonic crystal, QWP: quarter wave plate, LP: linear polarizer*

### 2- Analytical model of a pure electric coupling of light-to-BSW

Figure S2 illustrates the local coordinate frames $(x', y', z')$ of an incident plane wave, and $(x^L, y^L, z^L)$ and $(x^R, y^R, z^R)$ of the BSWs propagating on the left and right sides of the groove, respectively. It also show the global coordinate frame $(x, y, z)$ linked to the surface of the 1D photonic crystal. The incidence angle of the plane wave is $\beta$.

The model of a pure electric coupling between the incident light and the BSW suggests the projection of the amplitude of the incident field to the unit vectors $\vec{e} = (1,0,0)$ expressed in the local coordinate frames $(x^L, y^L, z^L)$ and $(x^R, y^R, z^R)$. These vectors are represented in red in Fig. S2.

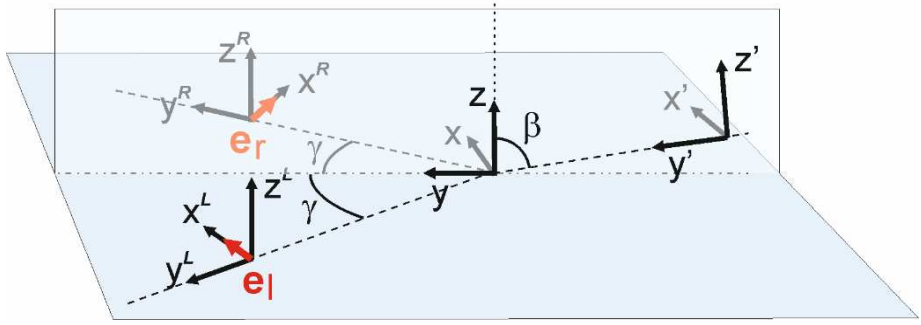

*Fig. S2. Schematics of the field projection and corresponding global and local coordinate frames*

Assigning $\gamma$ to be the angle between the groove and BSW trajectories, $\vec{e}_r$ and $\vec{e}_l$ expressed in $(x, y, z)$ read:

$$\vec{e}_r = \begin{bmatrix} \cos\gamma \\ -\sin\gamma \\ 0 \end{bmatrix} \tag{S1}$$

$$\vec{e}_l = \begin{bmatrix} \cos\gamma \\ \sin\gamma \\ 0 \end{bmatrix} \tag{S2}$$

To define the incident electric field $\vec{E}_{inc}$, we assume that the incident plane wave, initially polarized along $z'$ direction, is phase retarded by a quarter-wavelength along an axis tilted by an angle $\theta$ with respect to the $x'$-axis. We find, in the $(x, y, z)$ coordinate frame:

$$\vec{E}_{inc} \propto \begin{bmatrix} -(1+i)\sin\theta\cos\theta \\ (\cos^2\theta - i\sin^2\theta)\cos\beta \\ (\cos^2\theta - i\sin^2\theta)\sin\beta \end{bmatrix} \tag{S3}$$

Following the coupling model proposed in the article, the coupling rates $R_r$ and $R_l$ of the incident wave to the right and left BSWs, respectively, read:

$$R_r = \alpha|(1+i)\sin\theta\cos\theta\cos\gamma + (\cos^2\theta - i\sin^2\theta)\cos\beta\sin\gamma|^2, \tag{S4}$$

$$R_l = \alpha|(1+i)\sin\theta\cos\theta\cos\gamma - (\cos^2\theta - i\sin^2\theta)\cos\beta\sin\gamma|^2, \tag{S5}$$

where $\alpha$ is a constant. According to the phase matching condition (*i.e.*, linear momentum conservation) between the free space propagating incident wave and the BSWs, we have:

$$\frac{\omega}{c}\sin\beta = k_{BSW}\cos\gamma, \tag{S6}$$

where $k_{BSW}$ is the wave vector of the BSWs. In our case, $n_{eff} = 1.185$ and $\beta = 80°$, therefore we find $\gamma = 33.8°$.

Fig. S3 shows $R_r$ and $R_l$ as a function of the angle $\theta$. The blue and red curves represent the coupling rates to the left and right BSWs. These two coefficients, which describe a pure electric coupling in the BSW excitation process, show a $4\theta$ dependence. They are also shifted by an angle of about 8° 3- **Ellipticity of the magnetic field of an incident plane wave in the helicity plane of the TE-polarized BSWs.**

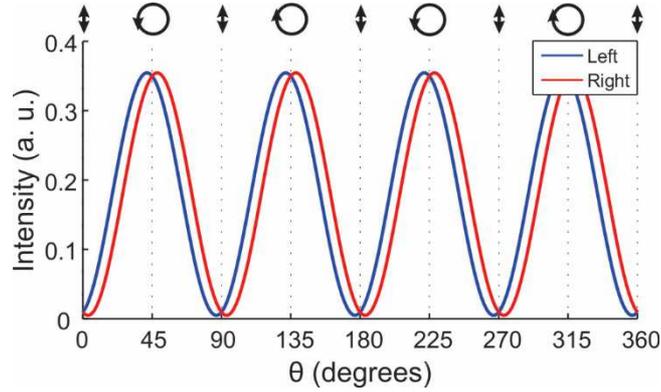

**Fig. S3.** *Plots of the coupling rates $R_r$ (red curve) and $R_l$ (blue curve) as a function of the angle $\theta$.*

According to the Maxwell-Faraday equation, the magnetic field of the incident plane wave can be expressed from Eq. S3 as:

$$\vec{H}_{inc} = \frac{\omega\varepsilon}{k_0}\begin{bmatrix} \cos^2\theta - i\sin^2\theta \\ (1+i)\sin\theta\cos\theta\cos\beta \\ (1+i)\sin\theta\cos\theta\sin\beta \end{bmatrix} \qquad (S7)$$

To examine the ellipticity of the incident magnetic field in the helicity planes of the BSW, one can make a transformation of $H_{inc}$ from the global coordinate frame $(x, y, z)$ to local ones $(x^L, y^L, z^L)$ and $(x^R, y^R, z^R)$. To this end, we define $\vec{H}_i = M_i \times \vec{H}_{inc}$, where $i = r, l$. $M_r$ and $M_l$ are the transformation matrices having the following forms for the right and left BSWs:

$$M_r = \begin{pmatrix} \cos\gamma & -\sin\gamma & 0 \\ \sin\gamma & \cos\gamma & 0 \\ 0 & 0 & 1 \end{pmatrix} \qquad (S81)$$

$$M_l = \begin{pmatrix} \cos\gamma & \sin\gamma & 0 \\ -\sin\gamma & \cos\gamma & 0 \\ 0 & 0 & 1 \end{pmatrix} \qquad (S2)$$

The ellipticity of the incident magnetic field is defined by the polar angle $(2\chi)$ of the Poincaré sphere. Figure S4 represents $2\chi$ in the helicity planes $(y^L, z^L)$ and $(y^R, z^R)$ of the left and right BSWs, respectively, as a function of $\theta$. The two curves show an oscillating behavior with a $2\theta$ dependence and an amplitude of 63.7°. They are in opposition from each other, i.e., shifted by 180°. Switching polarization handedness changes the sign of the angle $2\chi$, it is obviously helicity-dependent.

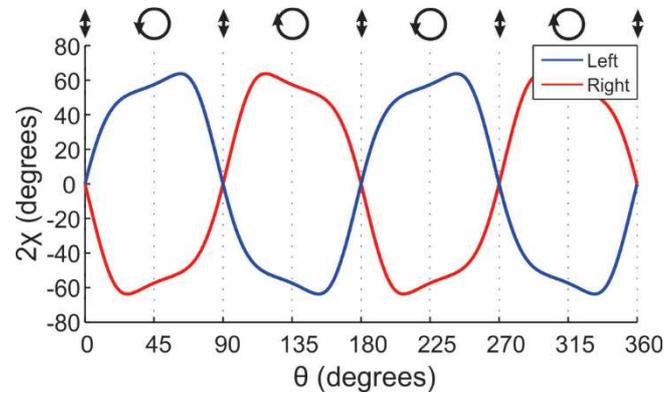

***Fig. S4.*** *Ellipticity $2\chi$ of the incident magnetic field in the helicity planes $(y^L, z^L)$ and $(y^R, z^R)$ of the left and right TE-polarized BSWs.*